# Novel Superstructure-Phase Two-Dimensional Material 1$T$-VSe$_2$ at High Pressure

Raimundas Sereika [1,2], Changyong Park [3], Curtis Kenney-Benson [3], Sateesh Bandaru [4,5], Niall J. English [5], Qiangwei Yin [6], Hechang Lei [6], Ning Chen [7], Cheng-Jun Sun [8], Steve M. Heald [8], Jichang Ren [9], Jun Chang [10], Yang Ding [1], Ho-kwang Mao [1,11]

[1]*Center for High Pressure Science and Technology Advanced Research, Beijing 100094, China*

[2]*Vytautas Magnus University, K. Donelaičio Str. 58, Kaunas 44248, Lithuania*

[3]*High Pressure Collaborative Access Team, X-ray Science Division, Argonne National Laboratory, Lemont, Illinois 60439, USA*

[4]*College of Materials and Environmental Engineering, Institute for Advanced Magnetic Materials, Hangzhou Dianzi University, Hangzhou 310018, China*

[5]*School of Chemical and Bioprocess Engineering, University College Dublin, Belfield, Dublin 4, Ireland*

[6]*Department of Physics and Beijing Key Laboratory of Opto-electronic Functional Materials & Micro-nano Devices, Renmin University of China, Beijing 100872, China*

[7]*Canadian Light Source, 44 Innovation Boulevard, Saskatoon, SK, S7N 2V3, Canada*

[8]*X-ray Science Division, Advanced Photon Source, Argonne National Laboratory, Lemont, Illinois 60439, USA*

[9]*Nano and Heterogeneous Materials Center, School of Materials Science and Engineering, Nanjing University of Science and Technology, Nanjing 210094, People's Republic of China.*

[10]*College of Physics and Information Technology, Shaanxi Normal University, Xi'an 710119, P. R. China*

[11]*Geophysical Laboratory, Carnegie Institution of Washington, Washington DC 20015, USA.*

*Corresponding authors: raimundas.sereika@hpstar.ac.cn; yang.ding@hpstar.ac.cn;

**Abstract**

A superstructure can elicit versatile new properties of materials by breaking their original geometrical symmetries. It is an important topic in the layered graphene-like two-dimensional transition-metal dichalcogenides (TMDs), but its origin remains unclear. Using diamond-anvil cell techniques, synchrotron x-ray diffraction, x-ray absorption, and the first-principles calculations, we show that the evolution from the weak Van der Waals bonding to the Heisenberg covalent bonding between layers induces an isostructural transition in quasi-two-dimensional 1$T$-type VSe$_2$ at high pressure. Furthermore, our results show that high-pressure induce a novel superstructure at 15.5 GPa, rather than suppress as it would normally, which is unexpected. It is driven by the Fermi surface nesting, enhanced by the pressure-induced distortion. The results suggest that the superstructure not only appears in the two-dimensional structure but also can emerge in the pressure-tuned three-dimensional structure with new symmetry and develop superconductivity.

Two-dimensional transition metal dichalcogenides (TMDs) exhibit rich physics, which provide an ideal 'playground' to study novel quantum states, as well as showing promising features for future advanced technological applications[1-3]. Indeed, various modulation-superstructures have been realized in TMDs[4,5]. Periodic lattice distortion (PLD) is often accompanied by charge density waves (CDW), raising the Coulombic and elastic energy as the consequence and being compensated by lowering the energy of occupied electronic states. According to Peierls instability, the distortion-induced kinetic energy gain is proportional to the non-interacting electronic susceptibility; Fermi-surface nesting would enhance the energy gain by sharply increasing the electronic susceptibility. However, in the TMDs with a superlattice, Density-Functional Theory (DFT) calculations have claimed that singularities in the susceptibility are not such a prerequisite towards realizing superstructures, while angle-resolved photoemission spectroscopy also found no sharp Fermi-surface nesting. Therefore, the origins of the TMD superstructures, which are commonly attributed to either electron-phonon coupling or Fermi-surface nesting, warrant further elucidation and clarification. Moreover, many studies have contributed to enhancing our understanding of low-temperature physics of TMDs. However, phenomena related to the superstructure transitions under high pressure have yet to be explored comprehensively in these 2-D systems, although recent progress in this regard has been somewhat encouraging[6-10].

Vanadium diselenide ($VSe_2$), in its 1*T* polytype, has a super-lattice incommensurate with the primitive lattice (*i.e.*, a case of periodic lattice distortion coupled to charge-density waves, PLD-CDW), which is distinct from those in the other 1*T* or 2*H* TMDs because of its CDW transition temperature increasing with pressure[11]. The PLD-CDW super-lattice in $VSe_2$ is characterized by a commensurate $4\mathbf{a_0} \times 4\mathbf{a_0}$ super-lattice forming in the layer plane and an incommensurate ~ $3c_0$ super-lattice forming perpendicularly to the layers at ~ 80 K[12]. The strong covalent bonding and small overlap of electron-wave functions inside the metallic layers result in quasi-two-dimensionality and the high anisotropy of physical properties at ambient conditions. The layers are linked with weak van der Waals forces allowing various creative materials-design possibilities, such as the intercalation of foreign atoms and molecules[13], exfoliation of the sample to the desired number of layers[14,15], as well as applications in electronic devices[16,17]. Intriguingly, recent studies of VSe$_2$ revealed dramatic changes in the CDW and new physical properties such as a pseudo-gap, Fermi arc, and emergent superconductivity[18-21]; naturally, this motivates us to investigate the behavior of this material under high pressure.

In this study, by using the diamond-anvil cell (DAC)[22], synchrotron x-ray diffraction (XRD), x-ray absorption (XAS), as well as the first-principles calculations, we discovered an isostructural transition at around 6.5 GPa and a novel superstructure PLD phase at ~ 15 GPa occurring in 1$T$-VSe$_2$. According to the XRD results and first-principles calculations, the high-pressure superstructure phase is associated with both the distortion of the structure and Fermi-surface nesting. The isostructural transitions in TMDs are commonly observed under high pressure[23,24]. However, it is unusual that application of high pressure induces a new symmetrized phase with a super-lattice, when pressure commonly suppresses such superstructures.

To investigate the superstructure and the phase transition at high pressure, we applied single-crystal zone-axis x-ray diffraction. This technique has been intensively used in transmission-electron microscopy (TEM) and has recently also been introduced into the field of synchrotron single-crystal x-ray diffraction[25,26]. The unique advantage of this technique lies in its ability to study very weak satellites in a tiny portion of the reciprocal space in the low scattering angle region. We measured the diffraction patterns along the [001] zone axis of 1$T$-VSe$_2$ at a variable pressure-temperature range (with details given in the Supporting Information). During compression, a super-lattice with commensurate peaks appeared along the original *a*-axis at ~15 GPa, 298 K (see Fig. 1). When lowering the temperature, we noticed that the transition temperature for the super-lattice decreased with increasing pressure within the experimental *P-T* range. However, single-crystal zone-axis x-ray diffraction is limited by its small reception of the reciprocal space, incomplete peak profiles, and less accurate lattice parameters[25,26]. Therefore, we also performed powder x-ray diffraction experiments to provide additional information on the larger reciprocal space in the high scattering angle region, more accurate lattice parameters, and better peak profiles.

In Figure 2a, we show the phase transitions observed by the synchrotron x-ray diffraction for 1*T*-VSe$_2$ powders. The structural transition occurs at 15.5 GPa, evidenced by a significant broadening of peak $(01\bar{1})$ and the appearance of new peaks in the diffraction patterns (Fig. 2b). In the meantime, the single-crystal zone-axis diffraction also shows some satellites appearing at (1/3, 0, 0) at nearly the same pressure value, thereby suggesting a 3×1×1 commensurate superstructure forming in a new symmetrized phase. By indexing the powder diffraction patterns of the new phase, we determined the sample transforms from trigonal $P\bar{3}m1$ into a monoclinic phase. Furthermore, powder x-ray diffraction measurements revealed an isostructural transition occurring around 6.7 GPa that is signaled by a sudden change in the pressure-dependent *c/a* ratio (see in Figs. 3b). A faster nonlinear decrease of *c/a* ratio with the pressure, before the transition, indicates that the cell parameter *c* is more compressible than *a* in this pressure range. Such anisotropic compressibility is attributed to the difference between the weaker van der Waals interlayer bonding and the stronger intralayer covalent bonding. Then, a slower, roughly linear decrease of *c/a*, observed after the transition, suggests that the cell parameters *a* and *c* have nearly equal compressibility. The interlayer space hedged by Se atoms shrinks continuously under pressure, where the shortest distance between Se atoms decreases before finally resulting in bond formation at the structural transition (Fig. 3d).

Determining a structure model directly from the powder x-ray diffraction patterns is challenging even at ambient conditions and becomes even more so in high-pressure environments where the background significantly complicates the intensity and profiles of the peaks. To tackle this problem, we created a 3×3×3 supercell based on the structure at the ambient conditions and searched for a possible high-pressure structural model using evolutionary metadynamics[27] implemented in the USPEX software[28]. Then, the resultant structure models were optimized using the Vienna *ab-initio* simulation package VASP[29] by minimizing the total energy and force on the atom with a converged criterion below $10^{-4}$ eV/Å (with more details given in Supporting Information). Three structure models were finally determined from the calculations, which all have super-lattices with the same symmetry of $C2/m$ (No. 12) but in different unit cell shapes: (1) for #1 model, $a = 11.7824$ Å, $b = 3.0502$ Å, $c = 12.0426$ Å, $\beta = 142.8996°$, $V = 261.07$ Å$^3$; (2) for #2 model, $a = 11.7824$ Å, $b = 3.0502$ Å, $c = 15.7368$ Å, $\beta = 152.5089°$, $V = 261.07$ Å$^3$; (3) for #3 model, $a = 15.9810$ Å, $b = 3.0744$ Å, $c = 5.3157$ Å, $\beta = 89.9929°$, $V = 261.17$ Å$^3$. These three models have similar total energies for their ground states at high pressure. To benchmark the results from the prediction, we simulated the diffraction patterns to compare them with the experimental patterns (with more details in Supporting Information). The generated powder XRD patterns from the three models all match well with that the experimental data (see Fig. 2c), but only the #3 structure model reproduces the experimental single-crystal zone-axis diffraction pattern (see inset of Fig. 2d and Fig. S3), while the other two models fail. In the zone-axis diffraction pattern of the high-pressure phase, the orientation of the 3×1×1 supercell relative to the lattice of the original ambient phase is $a_{super}||3a_{amb}$; $b_{super}||[110]_{amb}$; $c_{super}||c_{amb}$.

The bulk modulus of the sample is obtained by fitting the pressure-volume data to the third-order Birch-Murnaghan equation of state (Fig. 3a). For higher pressures, the increase of the bulk modulus with pressure indicates the hardening of the sample. The optimized structural model also reveals the origin of the 3×1×1 high-pressure superstructure (Figure 3c). For instance, the high-pressure superstructure shows a close similarity to the structure of the ambient phase, which can be derived from the ambient phase structure by displacing two V atoms (labeled as V2 in Figure 3c) out of the *a-b* plane with a small distance (~0.0026% along *c*-axis). Moreover, the octahedral site formed by one V2 atom and its surrounding 6 Se atoms is more distorted than other V-centered octahedral sites. It is the displacement of the V2 atom that breaks the original translation symmetry and results in a tripled periodicity along the ***a***-axis of the initial phase, as well as a new symmetrized structure. Such a superstructure manifests a structural distortion (in which V atoms deviate from their original positions and form a distorted octahedral site) which originates from the bonds forming between layers at high pressure.

In Figure 4, we plot the temperature-pressure dependent structural change from trigonal $P\bar{3}m1$ to monoclinic $C$ 2/m together with the CDW and superconductivity results from Ref. 21. Here, our data shows consistency with the CDW evolution, which is interrupted by the new structure at 12 GPa, 240 K. The fracture in the CDW onset curve at 4.5 GPa could be the same isostructural transition that we observed using powder XRD at 6.7 GPa (room temperature). The superconductivity appears after the CDW collapse in the superstructure at 15 GPa, 4 K.

To investigate how the transitions affect the electronic structures of the sample, we performed additional high-pressure XAS measurements at the Se *K*-edge (~12.66 keV) to explore the unoccupied *p*-bands of Se at high pressure. The room temperature X-ray absorption near edge structure (XANES) (demonstrated in Fig. 5) show significant pressure dependency changes in a data region A (~12.70 keV). It is characterized by two trends T1 and T2, representing the peak-energy position shift and the peak-intensity decrease with broadening during sample compression, respectively. The pressure dependency changes in energy region A (Fig. 5 and Figs. S5-S7) are approximately standard for the unoccupied DOS of Se *p* bands, which broaden significantly after the isostructural transition. The broadening of the occupied 4*p* bands is associated with the bonding formation (more delocalization) between the Se atoms in two different layers during the isostructural transition around 6.7 GPa. The isostructural transition eventually leads to the van der Waals force between layers evolving into the Heisenberg exchange interaction under pressure. In contrast, the transition at ~15.5 GPa causes no sizable changes in the peak A, implying that the second transition has no dramatic effects on the Se *p* bands. This is consistent with the XRD result that the superstructure is mainly associated with the tiny displacement of V atoms. However, according to our first-principles calculations, it is noticeable that for the ambient phase in symmetry $P\bar{3}m1$, the density of states (DOS) of vanadium at $E_F$ are dominant, while for the high-pressure phase in symmetry $C2/m$, the DOS of selenium is increased substantially – 'overwhelming' those of vanadium (Fig. S8). The change in the orbital components at Fermi level should be responsible for the continuous evolution of XANES feature in the energy region A observed in the XAS of the Se *K*-edge. Intriguingly, the exfoliation of other TMDs does not change peak A in a similar fashion in x-ray absorption's near-edge structure spectrum[30].

Furthermore, in low-dimensional systems, the formation of a super-lattice is usually associated with either structural distortion (*e.g.*, electron-phonon coupling) or Fermi-surface nesting. At ambient conditions, vanadium diselenide possesses a type II CDW with structural distortion (electron-phonon coupling)[18,31]. However, the high-pressure magneto-resistance and Hall measurements suggest successive electronic structural changes with Fermi-surface topology at 6 GPa and 12 GPa, which match relatively well our defined isostructural and super-lattice-type transitions, respectively. In addition, our calculations based on the high-pressure structure model #3 (displayed in Fig. 3c) also reveal a Fermi-surface nesting vector existing in the superstructure (Fig. S9). Considering that the structural distortion and the Fermi-surface nesting vector both exist in the new superstructure phase, the origin of the superstructure may be a Fermi-surface nesting driven periodic lattice distortion.

In conclusion, our experimental XRD and XAS room-temperature high-pressure studies revealed that 1$T$-VSe$_2$ undergoes two transitions: an isostructural one at 6.7 GPa and a structural one at 15.5 GPa. The first transition was associated with layer sliding and anisotropic-isotropic-contraction change between lattice parameters. At this transition, the weak van der Waals bonds between layers eventually transfers into the strong covalent bonds under pressure. The second transition induces a 3×1×1 superstructure, which has lower symmetry than $P\bar{3}m1$. Using theoretical structure-prediction tools, we found that the new phase should be monoclinic $C\,2/\mathrm{m}$ where the superstructure is associated with V atoms displacement. The first-principles calculations suggest that Fermi surface nesting is involved in the super-lattice formation because the high-pressure phase contains wave vectors of the electrons corresponding to the Fermi energy. This exciting (and, to our knowledge, unique) discovery proves that superstructure not only emerges in two-dimensional structures, but paves the way for pressure tuning of three-dimensional structures to manipulate and engineer novel symmetry for disparate real-world applications, such as developing superconductivity.

**Acknowledgements**

Portions of this work were performed at HPCAT (Sector 16), and XSD (Sector 20) of Advanced Photon Source (APS), Argonne National Laboratory. HPCAT operations are supported by DOE-NNSA's Office of Experimental Sciences. XSD operations are supported by the U.S. Department of Energy (DOE) and the Canadian Light Source (CLS). The APS is a DOE Office of Science User Facility operated for the DOE Office of Science by Argonne National Laboratory under Contract No. DE-AC02-06CH11357. Y.D and H.-k.M. acknowledges the support from National Key Research and Development Program of China 2018YFA0305703, Science Challenge Project, No TZ2016001 and The National Natural Science Foundation of China (NSFC): U1930401, 11874075. H.C.L. acknowledges the support from the National Key Research and Development Program of China (Grants No. 2016YFA0300504), the NSFC (No. 11574394, 11774423, 11822412), the Fundamental Research Funds for the Central Universities, and the Research Funds of Renmin University of China (RUC) (15XNLQ07, 18XNLG14, 19XNLG17). S.B. and N.J.E. thank Science Foundation Ireland for support under the SFI-NSFC bilateral programme (SFI 17/NSFC/5229).

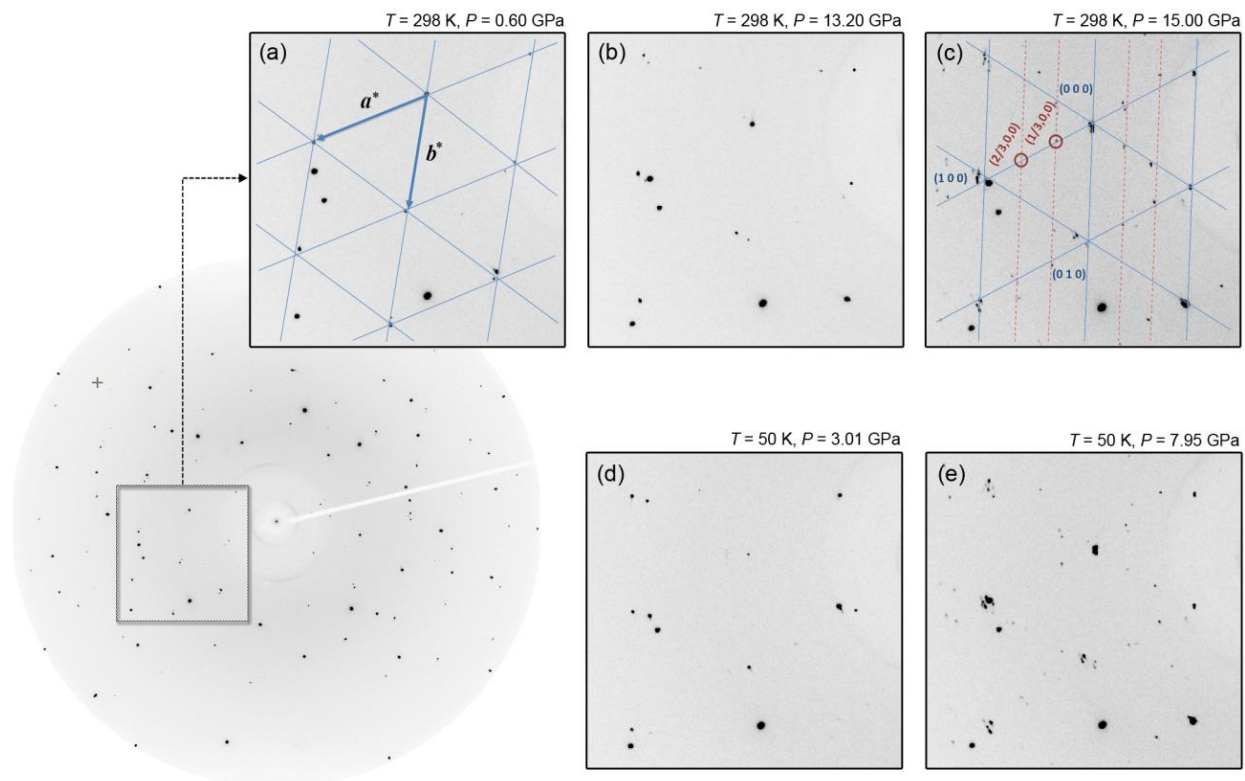

**Figure 1.** Diffraction images at different temperature-pressure conditions indicating a phase transition with superlattice reflections in 1$T$-VSe$_2$. New peaks appear on the ***a***-axis, as marked by dashed red lines, prior to the ambient lattice, which is denoted by the solid blue lines (unmarked images given in the Supporting Information). Here, $a^*$ and $b^*$ are the trigonal lattice parameters in the reciprocal space before the transition. Measurement data also revealed crystal twinning, which is evident by the presence of the additional symmetrical reflections. Note: this XRD measurement is not intended to show the CDW. Images may contain peaks from diamonds, Ne gas, and the cryostat window.

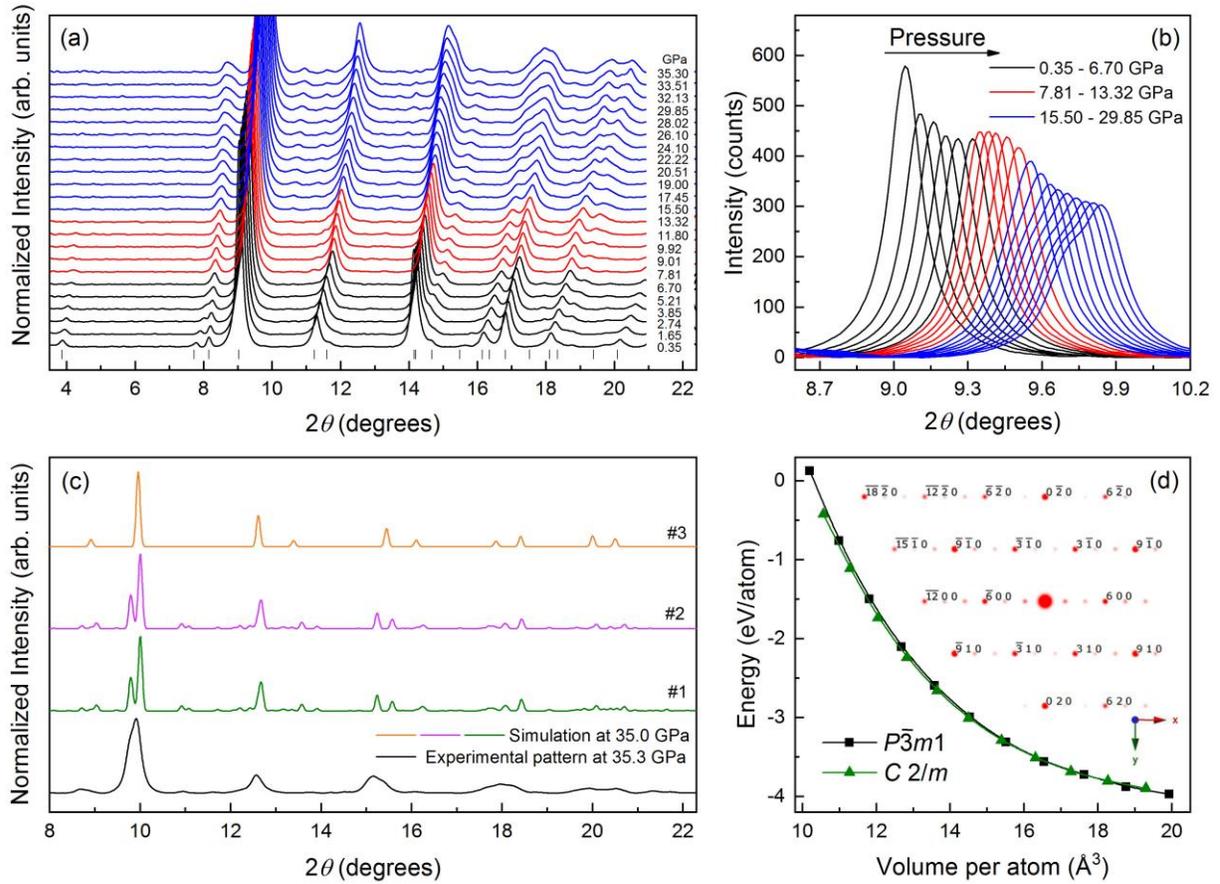

**Figure 2.** Synchrotron x-ray diffraction data for $1T$-$VSe_2$ powders compared with the simulated results. The X-ray wavelength is 0.4133 Å. (a) The evolution of diffraction patterns over sample compression at room temperature. The positions of the initial $P\bar{3}m1$ Bragg reflections are marked by vertical bars. (b) The profile of the $(01\bar{1})$ peak at different ranges of pressure indicating its continuous intensity reduction and splitting. (c) The simulated XRD patterns for the high-pressure phase compared with experimental data at the same pressure. (d) Energies as a function of volume/atom calculated using GGA+$U_{eff}$. The inset shows a portion of the simulated reciprocal lattice of the monoclinic phase #3 (full image is given in the Supporting Information).

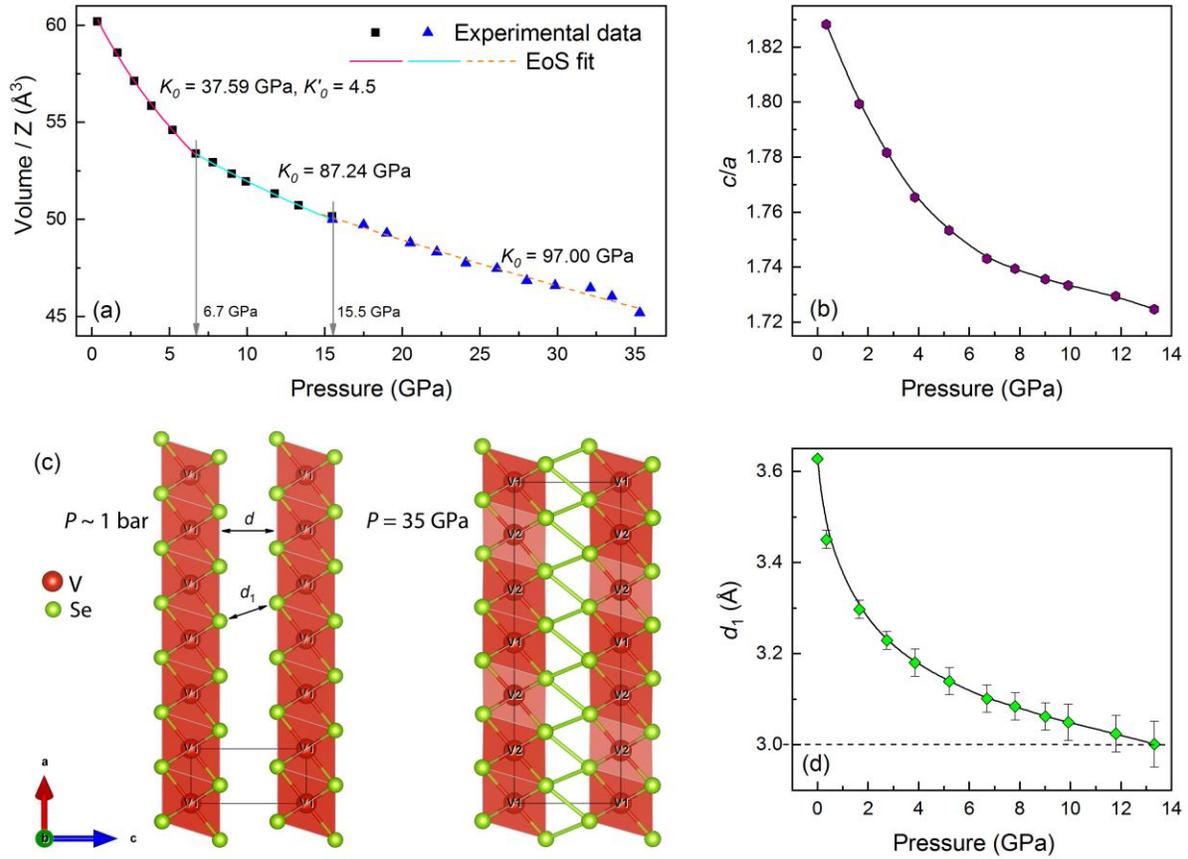

**Figure 3.** Changes of the unit cell in the 1$T$-VSe$_2$ at different pressures. (a) Volumes per formula unit as a function of pressure for $P\bar{3}m1$ (black squares) and $C\,2/m$ (blue triangles) phases. The solid and dashed lines are the calculated third-order Birch−Murnaghan equation of state (EoS) fit to the experimental data. (b) Pressure dependence of the axial ratio $c/a$ before the structural transition. (c) Crystal structure view of the ***a*-*c*** plane in the ambient $P\bar{3}m1$ phase and simulated high-pressure $C\,2/m$ phase (35 GPa pressure). Thick solid lines indicate the unit cell. 2-D layers consist of vanadium-centered polyhedrons stacked along the ***c***-axis. (d) Pressure dependence of the shortest Se-Se distance $d_1$ in the interlayer space.

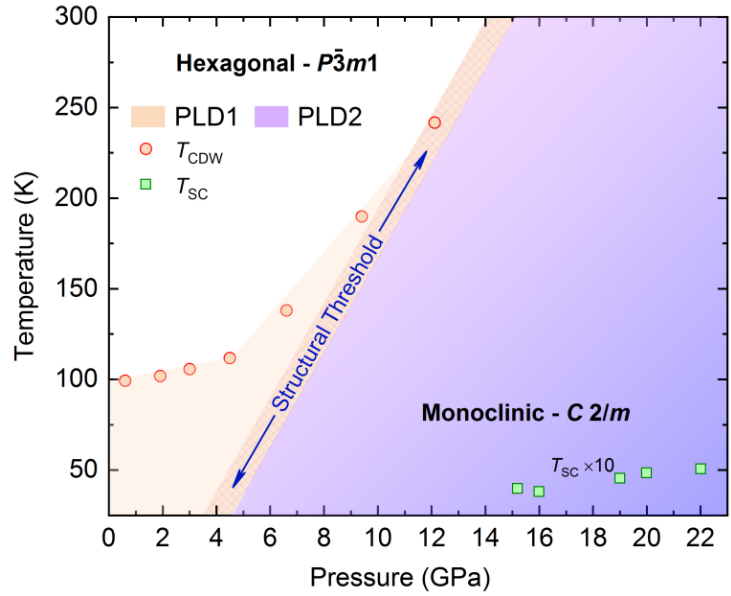

**Figure 4.** Structural changes in the 1$T$-VSe$_2$ at different temperatures and pressures. Periodic lattice distortions in the pressure range from 0 to 23 GPa and a temperature range from 25 to 300 K for 1$T$-VSe$_2$. The structural threshold is drawn from our low temperature single-crystal zone-axis XRD data. The data points for $T_{CDW}$ and $T_{SC}$ were taken from Ref. 21.

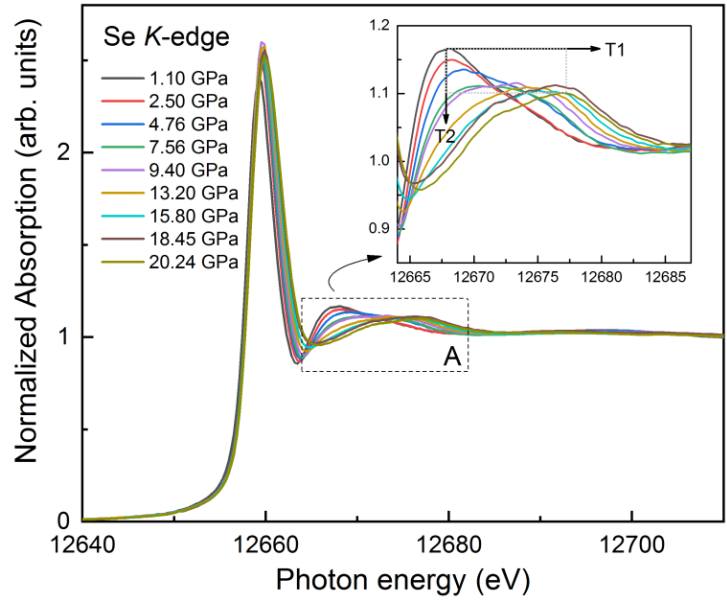

**Figure 5.** Normalized absorption spectra of 1$T$-VSe$_2$ at the Se $K$-edge as a function of pressure. The inset shows zoomed area "A" where two trends T1 and T2 represent pressure induced progressive feature peak position drifting and feature peak intensity decrease, respectively.